\documentstyle[12pt]{article}

\def\s2{\frac{1}{\sqrt2}}

\def\beq{\begin{equation}}
\def\eeq{\end{equation}}
\def\beqa{\begin{eqnarray}}
\def\eeqa{\end{eqnarray}}
\def\IZ{\relax\ifmmode\hbox{\ss Z\kern-.4em Z}\else{\ss Z\kern-.4em
Z}\fi}
\def\IP{\relax{\rm I\kern-.18em P}}
\def\IC{\relax\hbox{\kern.25em$\inbar\kern-.3em{\rm C}$}}

\def\cp#1{\relax\ifmmode {\IP\kern-2pt{}_{#1}}\else
$\IP\kern-2pt{}_{#1}$\fi}

\def\Tr{\mathop{\rm Tr}}

\def\be{\begin{equation}}
\def\ee{\end{equation}}

\def\WW{W^\alpha W_\alpha}
\def\lbarl{\lambda^\alpha \lambda_\alpha}
\def\IZ{{\bf Z}}
\def\Avg#1{\left\langle #1 \right\rangle}

\def\d{\partial}

\topmargin
-2cm
\textwidth
15cm
\textheight
24cm
\oddsidemargin
1cm
\evensidemargin
1cm
\begin{document}
\date{}

\title{ Gaugino Condensation, Duality and Supersymmetry Breaking}

\author{Fernando  Quevedo \thanks{
Contribution to the Conference on S-duality and Mirror
Symmetry, Trieste, June 1995}\\[4mm]
\normalsize  Theory Division CERN \\
\normalsize CH-1211 Geneva 23 \\
\normalsize Switzerland.\\
\normalsize e-mail:quevedo@nxth04.cern.ch }

\maketitle
\vspace{-4in}
\rightline{CERN-TH/95-308}
\rightline{hep--th/9511131}
\vspace{3.8in}

\begin{abstract}

The status of gaugino
condensation  in low-energy string theory is reviewed.
Emphasis is given to the determination of the
efective action below condensation scale in terms
of the 2PI and Wilson  actions.
 We illustrate how the different perturbative duality symmetries
survive
 this simple nonperturbative phenomenon, providing evidence
for the believe that these are exact nonperturbative symmetries of
string theory.
Consistency with $T$ duality lifts the moduli
degeneracy. The $B_{\mu\nu}-axion$ duality also survives in a
nontrivial  way in which the degree of freedom corresponding to
$B_{\mu\nu}$ is replaced by a massive $H_{\mu\nu\rho}$ field
but duality is preserved.
 $S$ duality may also  be implemented in  this process.
Some general problems of this
mechanism
are mentioned   and
the possible nonperturbative scenarios for supersymmetry breaking
in string  theory are discussed.

\end{abstract}
\vskip 1in

\begin{flushleft}
CERN-TH/95-308 \\
November 1995 \\
\end{flushleft}
\maketitle

\newpage

\section{Introduction}

In the efforts to extract a relation between string theory
and physics, we find two main problems, namely how
 the large vacuum degeneracy is lifted and how
 supersymmetry is broken at low energies.
These problems, when present at string tree level,
cannot be solved at any order in string  perturbation
theory. The reason is the following:
It is known that at tree-level, setting all the matter
fields to zero forces the superpotential to vanish,
for {\it any} value of the moduli and dilaton fields.
The corresponding scalar
potential vanishes implying flat directions for the moduli and
dilaton.
Also, the $F$ and $D$ auxiliary fields, which
are the order parameters for supersymmetry breaking,
 vanish  in this situation, implying unbroken
supersymmetry.
Since the
superpotential does not get renormalized in perturbation theory,
if it vanish at tree level it will also vanish
at all orders of string perturbation theory.
Then the F-term part of the potential also vanishes
perturbatively.
The only perturbative correction that could alter
this situation is the generation of a Fayet-Iliopoulos
D-term by an `anomalous' $U(1)$, usually present in 4D strings.
However, in all the cases considered so far there are
charged fields getting  nonvanishing vev's which cancel the
$D$-term,  breaking  gauge symmetries instead of
supersymmetry.

Therefore these problems are exact in perturbation theory and the
only
hope to solve them  is nonperturbative physics.
This has a good  and a bad side. The good side is that
nonperturbative effects represent the most natural way to
generate large hierarchies due to their exponential
suppression, this is precisely what is needed to obtain the
Weinberg-Salam scale from the fundamental string or Planck scale.
The bad side is that despite many efforts, we do not yet have
 a nonperturbative formulation of string theory.
At the moment, the only concrete nonperturbative information
we can extract is from the purely {\it field theoretical}
 nonperturbative
effects inside string theory. Probably the simplest and
certainly the most studied of those effects is gaugino
condensation in a hidden sector of the gauge group, since it
has the potential of breaking supersymmetry as well as lifting
some of the flat directions, as we will presently discuss.

\section{Gaugino Condensation}

The idea of breaking supersymmetry in a dynamical way was
first presented in refs.~\cite{witten}. In those articles a
general topological argument was developed in terms of the
Witten index $Tr(-)^F$, showing that dynamical supersymmetry
breaking {\it cannot} be achieved unless there is chiral matter or
we include supergravity effects for which the index
argument  does not apply.
This was subsequently verified by explicitly studying gaugino
condensation in pure supersymmetric Yang-Mills, a vector-like
theory, for which gauginos condense but do not break
global supersymmetry \cite{vy} (for a review see \cite{amati}).
Breaking  global supersymmetry with chiral matter
was an open possibility in principle, but this approach ran into
many problems when tried to be realized in practice.

The situation improved very much with the coupling to supergravity.
 The reason was that simple
gaugino condensation was argued to be sufficient to break
supersymmetry once the coupling to gravity was included. This
 works in a hidden sector mechanism where gravity is the messenger
of supersymmetry breaking to the observable sector
\cite{peter}.
Furthermore,  string theory provided a natural realization of this
mechanism \cite{din,drsw}\ by having
naturally a hidden sector especially in the $E_8\times E_8$ versions.
Also, it gave another direction to the mechanism by the fact that
gauge couplings are field dependent
(as anticipated  for supergravity models in ref.~\cite{fgn}). This
same
fact raised the hope that gaugino condensation could lift the
moduli and dilaton  flat directions, but soon it was recognized that
it only changed flat to runaway potentials, thus destabilizing those
fields in the `wrong' direction (zero gauge coupling and infinite
radius)\footnote{The possibility of a nonvanishing
$\Avg{H_{ijk}}$ stabilizing the potential with vanishing
cosmological constant \cite{drsw}, was discarded after it was
realized that this
field was always quantized, breaking supersymmetry at the Planck
scale,
also its incorporation does not seem consistent with $T$-duality.}.

A simple way to see this is by setting the gaugino condensate
$\Avg{\lbarl}\sim \Lambda^3$ with $\Lambda\sim M\exp(-1/(bg^2))$,
 the renormalization
group invariant scale. Here $M\sim 10^{19}$ Gev is the
compactification
scale, $b$ the coefficient of the one-loop beta function of the
hidden sector group and
$g$ the corresponding gauge coupling. In string theory we have that
$4\pi g^{-2}\sim \Avg{S+S^*}$ where $S$ is the chiral dilaton field
(including also the axion and fermionic partner). Also,
$M^{-1}\sim \Avg{T+T^*}$ with $T$ being one of the moduli
fields. Substituting naively $\Avg{\lbarl}$ into the lagrangian
induces a scalar
potential for the real parts of $S$ and $T$ ($S_R$ and $T_R$
respectively), namely $V(S_R,T_R)\sim
\frac{1}{S_RT_R^3}\exp(-3S_R/4\pi b)$.
This potential has a runaway behaviour for both $S_R$ and $T_R$,
as advertized.

 The $T$ dependence of the potential was completely changed after the
consideration of target space or $T$ duality. In its simplest form,
this symmetry acts on the field $T$ as an $SL(2,\IZ)$
symmetry:
\be
T\rightarrow \frac{a\, T-i\, b}{i\, c\,T+d}, \qquad a\, d-b\, c=1.
\ee
 It was shown \cite{filq}, that imposing this symmetry
changes the structure of the scalar potential for the moduli fields
in such a way that it develops a minimum at $T\sim 1.2$ (in
string units), whereas the potential blows-up at the
decompactification
limit ($T_R\rightarrow\infty$), as desired.
The modifications due to imposing $T$ duality can be traced to the
fact
that the gauge couplings get moduli dependent threshold corrections
from
loops of heavy string states \cite{dkl}. This in turn generates a
moduli dependence on the superpotential induced by gaugino
condensation
of the form $W(S,T)\sim\eta(T)^{-6}\exp(-3S/8\pi b)$ with
$\eta(T)$ the Dedekind function.

This mechanism however did not help in changing the runaway behaviour
of the potential in the direction of $S$. For stabilizing $S$, the
only
proposal was to consider gaugino condensation of a nonsemisimple
 gauge group, inducing a sum of exponentials in the superpotential
$W(S)\sim \sum_i{\alpha_i \exp(-3S/8\pi b_i)}$
which conspire to generate a local minimum for $S$ \cite{krasnikov}.
These have been named `racetrack' models in the recent literature.

It was later found that combining the previous ideas together with
the addition of  matter fields in the hidden sector (natural in many
string models)\cite{lt,ccm},  was sufficient to find a minimum with
almost all
the right properties, namely, $S$ and $T$ fixed at the desired value,
$S_R\sim 25, T_R\sim 1$, supersymmetry broken at a small
scale ($\sim 10^{2-4}$ GeV) in the observable sector, etc.
This lead to studies of the induced soft breaking terms at low
energies.

Besides that relative succes, there are at least five problems that
assures us that we are far from a satisfactory treatment of
these issues.

\begin{description}
\item[(i)] Unlike the case for $T$, fixing  the
$vev$ of the dilaton field $S$, at the phenomenologically
interesting value, is not achieved in a satisfactory way.
The conspiracy of several condensates with hidden
matter to generate a local minimum
at a good value, requires certain amount of fine tunning and
cannot be called natural.
\item[(ii)] The  cosmological constant turns out to be
always negative,
which looks like an unsourmountable problem at present. This
also makes the analysis of soft breaking terms less reliable,
because in order to talk about them, a constant
piece has to be added to the lagrangian   that
cancels the cosmological constant. It is then hard to
believe that the unknown mechanism generating this term would
leave the results on  soft breaking terms (such as
small gaugino masses) untouched.
\item[(iii)] The derivation of the effective theory below
condensation is not completely understood. There are several
approaches to this and the exact relation among them is not
completely
clear.
\item[(iv)] There is an inherently stringy problem which is due to
the fact that the $S$ field is not stringy. $S$ is only the dual
of another field, $L$ which is the one created by string vertex
operators, having the dilaton and the antisymmetric tensor field
$B_{\mu\nu}$ (instead of the axion) as  the bosonic components.
The problem resides in the fact that, if there is not a Peccei-Quinn
(PQ)
symmetry $S\rightarrow S+i\, constant$, as in the many condensates
scenario,
it is not clear if the theory in terms of $S$ is any longer dual to
the $L$ theory. This sets serious doubts on whether the $S$ approach
mentioned above is valid at all. Another way to express this problem
is to ask if it is possible to formulate directly gaugino
condensation in
terms of the stringy field $L$.
\item[(v)] Finally, even if the previous problems were solved, there
are
at least two serious cosmological problems for the gaugino
condensation
scenario. First, it was found under very general grounds, that
it was not possible to get inflation with the type of dilaton
potentials
obtained from gaugino condensation \cite{bs}. Second is the so-called
`cosmological moduli problem' which applies to any (nonrenormalizble)
hidden sector scenario including gaugino condensation
\cite{modprob}. In this case,
it can be shown that the moduli and dilaton fields acquire masses
of the electroweak scale ($\sim 10^2$ GeV) after supersymmetry
breaking.
Therefore if  stable, they overclose the universe, if
 unstable, they  destroy  nucleosynthesis by their
late decay, since they only have gravitational strength interactions.
\end{description}

In the next section, I will present a general description
of the effective theory below condensation scale, addressing the
issue
of problem (iii) above. Section 4 will show the solution of
problem (iv) whereas in section 5, I will discuss ideas towards
solving
problems (i) and (v). The resolution of problem (ii) is
left to the reader.

\section{Wilson vs 2PI Actions}

 To study the effects of gaugino condensation we should be
able to answer the following questions:
Do gauginos condense? If so, is supersymmetry broken by this
effect? What is the effective theory below the scale
of condensation?
In order to answer these questions, several ideas have been put
forward
\cite{vy,fgn,drsw,mr}. Let me revise briefly the different
approaches.

In ref.~\cite{vy}, a chiral superfield $U$ was introduced
representing the condensate $\WW$. The effective supersymmetric
theory
in terms of $U$ was found by matching the anomaly of an
original $R$-symmetry of the underlying
supersymmetric Yang-Mills action.

In refs.~\cite{drsw}, \cite{kp},
the same anomalous symmetry was used
to reproduce the effective action below condensation scale,
without the need of introducing $U$. That gave rise
to the superpotential $W(S)\sim\exp(-3S/8\pi b)$ mentioned
before. The earlier approach of ref.~\cite{fgn} was
based on the direct substitution of $\lbarl$ in
the original supergravity lagrangian. A more recent
analysis of ref.~\cite{mr}, uses a Nambu-Jona-Laisinio
approach to describe the condensation mechanism.

 Even though some of these approaches
gave similar results, there are important differences among them.
In particular, following ref.~\cite{fgn}, since they
substitute $\lbarl$ directly into the supersymmetric action
in components, the effective lagrangian is not explicitly
supersymmetric unlike for instance the results of
ref.~\cite{drsw}. \footnote{These two
approaches were shown to be equivalent in
ref.~\cite{kl}, once the
superconformal structure of the original supergravity action
is considered in detail, giving rise an explicit supersymmetric
action as in \cite{drsw}}
 Also, the approach of
\cite{mr}, even though it reproduces  the
results in \cite{vy} at tree-level, by including quantum
corrections, they find very different results, for instance,
the dilaton could be stabilized with a single condensing group.
Finally the formalisms of \cite{vy} and \cite{drsw}
have been compared in \cite{lt,kl}. They eliminate
the field $U$ by {\it assuming} it does not break
global supersymmetry, {\it ie} by using $\partial W/
\partial U=0$ and find agreement between the two methods.
However this condition should not be imposed beforehand and it
is not well justified in the supergravity case.

We can see there is no satisfactory understanding of the effective
theory below condensation. Furthermore, the
anomalous symmetry argument which is the most solid
description of the single condensing case, cannot be  used for the
interesting case of several condensing groups.

We will now present a self contained discussion which
will at the end identify the main approaches with
known field theory quantities, {\it ie} the 2PI and
Wilsonian effective actions \cite{bdqq2}, and mention
how these two approaches
are actually related in a consistent manner.

\subsection{Supergravity Basics}

Since the
fields $S$ and $T$ are expected to have very large vev's,
it is more convenient to work with local supersymmetry without
 taking the Planck scale to $\infty$.
The most general action for chiral matter
supermultiplets $\Sigma$ coupled to supergravity
can  be written as   \cite{cfgvp}:
\begin{eqnarray}
{\cal I}&=&\int d^4x \; \left\{ -\frac{3}{4}
[S_0 S_0^* e^{-K(\Sigma,\Sigma^*)
/3}]_D+\right. \\
 & &\left.[S_0^3 W(\Sigma)]_F + [\frac{1}{4}f_{ab}(\Sigma)
W^{\alpha a} W^{b}_{\alpha} ]_F + {\rm cc}\right\}\nonumber
\end{eqnarray}
where  the K\"ahler potential $K(\Sigma,\Sigma^*)$, the
superpotential
$W(\Sigma)$
and  the gauge kinetic function $f_{ab}(\Sigma)$ define a particular
theory. The field $S_0$ is an extra chiral superfield
called `the compensator'. Its existence is due to the fact that
action (2) is not only invariant under super Poincar\'e symmetries
but
 under the full superconformal symmetry.
This simplifies the treatment of the theory in particular
the calculation of the action in components. Super Poincar\'e
supergravity
is easily obtained by explicitly fixing the
field $S_0$ to a particular value, it is usually chosen in such a way
that the coefficient of the Einstein term in the action is just
Newton's constant.

Two symmetries of the superconformal algebra have a particular
importance
for us: Weyl and chiral $U(1)$ transformations. These two symmetries
 do not commute
with  supersymmetry. The chiral $U(1)$ group
is at the origin of the R-symmetry of Poincar\'e theories.  Weyl
and chiral transformations with parameters $\lambda$
and $\theta$ respectively,
act on component fields with a factor
$e^{w_j\lambda + in_j\theta/2}$,
$w_j$ and $n_j$ being the Weyl and chiral weights of the component
field.
For a left-handed chiral multiplet $(z,\psi,f)$, one finds the
following
weights:
\begin{eqnarray}
z : & w,  & n=w , \nonumber \\
\psi :& w+{1\over2}, & n-{3\over2}, \nonumber \\
f :& w+1, &  n-3.
\end{eqnarray}
Chiral matter multiplets $\Sigma$ have $w=n=0$, except for $S_0$
which has
$w=n=1$. The chiral multiplet of gauge field strength $W^a$ has
$w=n=3/2$.
The $U(1)$ transformations of (left-handed)
gauginos and chiral fermions are therefore:
\be
\lambda^a \longrightarrow e^{3i\theta/4}\lambda^a ,\qquad\qquad
\psi \longrightarrow e^{-3i\theta/4}\psi.
\ee
These transformations generate a gauge-chiral
$U(1)$ mixed anomaly. This anomaly can be cancelled
by the `Green-Schwarz' counterterm \cite{dfkz,kl}\ :
\be
\Delta {\cal I}= - c \left\{\int d^4x \; [
{1\over4}\,Tr \WW\, \log S_0]_F+{\rm cc}\, \right\} .
\ee
where
$
c =  {3 \over 2 \pi } \left[ C(G) - \sum_I C\left(R_I \right)
\right],
$
(C here represents the Casimir of the representation,
for the case without matter we have that $c=8\pi b$).
This counterterm is claimed to cancel the anomaly to all orders in
perturbation
theory \cite{kl}\ and plays an important role
in what follows.

The action (12) has also a
 symmetry under K\"ahler transformations:
$K\rightarrow K+\varphi(\Sigma)+\varphi^*(\Sigma^*),
\, W\rightarrow e^{-\varphi(\Sigma)}\,  W $
since any such a transformation can be absorbed by redefining $S_0$:
$S_0\longrightarrow e^{\varphi/3}S_0$.

\subsection{The Wilson Effective Action}

Let us now restrict to a simple case that has all the properties we
need
to discuss gaugino condensation, {\it ie} a single
chiral multiplet $S$ coupled to supergravity and a nonabelian
gauge group with $K=K_p(S+S^*)$ arbitrary, $W(S)=0$
and $f(S)=S$. This  is the case for the dilaton in
string theory at the perturbative level. This defines the
effective (Wilson) action at scales $M\geq E\geq \Lambda$.
We are interested in the Wilson action at
scales $\Lambda\geq E\geq 10^2$ GeV in which we
expect that gauginos have condensed and $S$ is the only
degree of freedom, that means we want to integrate out
the full gauge supermultiplet to obtain the
effective action for $S$ at low energies.
This is precisely the approach of ref. \cite{drsw}\ mentioned above.
We need to compute:
\begin{eqnarray}
e^{i\Gamma(S,S_0)}&\equiv
&\int DV \exp{i\int d^4x}\left\{ [\left(S- c\,
\log{S_0}\right)\right. \nonumber   \\
 & &
 \left. \Tr\, \WW ]_F +{\rm cc}\right\}
\end{eqnarray}
First of all we can observe that $\Gamma(S,S_0)$ depends on its
arguments
only through the combination $S_0\exp(-S/c)$.
Second, since the result of the integration has to be superconformal
invariant (because the anomaly is cancelled), we
know that $\Gamma [S_0\exp(-S/c)]$ has to be
written in the form of equation (2)
(plus higher derivative terms)
with $f=0$ since there
are no gauge fields. Since the powers of $S_0$ are
exactly given by (2) and $S_0$ only appears multiplying
$\exp(-S/c)$ we can just read the super and K\"ahler potentials
to be:
\begin{eqnarray}
W(S)&=&w e^{-3S/c} \nonumber \\
e^{-K/3}&=&e^{-K_p/3}-k\,e^{-(S+S^*)/c}
\end{eqnarray}
where $w$ and $k$ are arbitrary constants ($k>0$
to assure positive kinetic energy).
The superpotential is just the one found in \cite{drsw}. The
correction to the K\"ahler potential is new
\cite{bdqq2}. Notice that both
are corrections of  order $\exp{-1/g^2}$ as expected.
A word of caution is in order.
 Unlike the superpotential which has
no corrections in perturbation theory, the K\"ahler
potential can be corrected order by order in perturbation
theory, therefore in practice the perturbative part
of the K\"ahler potential $K_p$ is simply unknown and
for weak coupling those corrections are bigger than
the nonperturbative correction found here. Our result
could be useful, only after the exact perturbative
K\"ahler potential is known. It is still interesting to
realize that such a simple symmetry argument can give us
the exact expressions for the {\it nonperturbative} super and
K\"ahler
potentials, without the need of holomorphy!

\subsection{The 2PI Effective Action}

To answer the questions posed at the beginning
of this chapter, {\it ie} whether
gauginos condense and break supersymmetry, it
is convenient to think
about the case of spontaneous breaking of gauge symmetries.
In that case we minimize the effective potential for
a Higgs field, obtained
from the 1PI effective action and see if the
minimum breaks or not the corresponding gauge symmetry.
In our case, we are interested in the expectation value
of a composite field, namely $\lbarl$ or its
supersymmetric expression $\WW$. Therefore we need
the so-called two particle irreducible effective
action.

We start then with the generating functional in
the presence of an external current $J$ coupled to
the operator that we want the expectation value of,
namely,
$\WW$:
\begin{eqnarray}
e^{i{\cal W}[S,S_0,J]}&\equiv &\int DV \exp i\int d^4x\left\{
[(S-c\log S_0 \right. \nonumber \\
  & & \left.  + J ) \Tr\WW]_F +{\rm cc}\right\}
\end{eqnarray}
{}From this we have
\be
\frac{\delta{\cal W}}{\delta J}=\Avg\WW\equiv U
 \ee
and define the 2PI action as
\be
\Gamma[S,S_0,\hat U]\equiv {\cal W}-\int d^4x\left(\hat U J\right)
\ee
To find the explicit form of $\Gamma$ we
use the fact that ${\cal W} $ depends on its three arguments
only thorugh the combination $S+J-c\log S_0$,
therefore, we can see that $\delta \Gamma/\delta (S-c\log S_0)
=\delta \Gamma/\delta J=\hat U$. Integrating this
equation
determines the dependence of $\Gamma$ in $S$ and
$S_0$:
\be
\Gamma[S,S_0,\hat U]=\hat U\left(S-c\log S_0\right)+\Xi (\hat U)
\ee
where $\Xi (\hat U)$ can be determined using symmetry arguments
as follows.
First we define a chiral superfield $U$ by
$\hat U\equiv S_0^3 U$. Therefore $U$ is a standard chiral superfield
with vanishing chiral and conformal weight ($w=n=0$).
Then $\Gamma[S, S_0,U S_0^3]$ can be writeen in the form (2)
with chiral fields $S$ and $U$.
Again the fact that the $S_0$ dependence of (2) is
very restricted, allows us to just read again the
corresponding K\"ahler and superpotential.
We find:
\begin{eqnarray}
W[S,U]&=&U[S+\frac{c}{3}\log U+\xi]\nonumber\\
e^{-K/3}&=&e^{-K_p/3}-a\, \left(UU^*\right)^{1/3}
\end{eqnarray}
Here $\xi$ is an arbitrary constant.
We can see that the superpotential corresponds to
the one found in \cite{vy}. The K\"ahler potential
is new, in \cite{vy}\ it was  found  for the global case, to which
this
reduces in the global limit.

Notice that we have identified the two main approaches to gaugino
condensation with the two relevant actions in field theory,
namely the Wilson and 2PI effective actions.
Our approach to the 2PI action is a reinterpretation of the
one in \cite{vy}. We have to stress that in our treatment
$U$ is only a {\it classical} field, not to be integrated out in any
path integral. It also does not make sense to consider loop
corrections to its potential, this solves the question
raised in \cite{mr} where loop corrections to the
$U$ potential could change the tree level results.
Furthermore, since $U$ is classical we can eliminate it by
just solving its field equations:$\d\Gamma/\d U=0$.
(Since this implies $J=0$, it makes equations (11) and (9)
reduce to (7).) These equations cannot
be solved explicitly but we find the solution in
an $1/\Lambda$ expansion. We find that the solution of these
equations reproduce the Wilson action derived in the
previous subsection (obtaining both $W(S)$ and $K(S+S^*)$
as in equation (9)) plus extra terms suppressed by inverse powers
of the condensation scale. This shows explicitly the relation
between the two approaches.

We can also consider the case of several condensates.
This case shows the power of the techniques used previously.
Following the original discussions of \cite{drsw}\
it was needed to use the PQ symmetry
of $S$ to cancel the $U(1)_R$ anomaly, however
when there are several condensing groups we would have neede several
$S$ fields
to cancel the anomaly (see \cite{bdqq2}) but there is only
one $S$ field in string theory. In our approach
however, we use the counterterm (5) which in the case
of several groups is a sum of terms \cite{kl}. Therefore
we have one counterterm for each group and so the path integrals
just factorize into products for each of the {\it many}
condensates, implying that the total superpotential
($W$) and $e^{-K/3}$ functions are the sum
of the ones for one single condensate. This is
the first real {\it derivation} of this well used
result!

By studying the effective potential for $U$ we recover the previously
known results. For one condensate and field independent
gauge couplings (no field $S$) the gauginos
condense ($U\neq 0$) but supersymmetry is unbroken.
For field dependendt gauge coupling, the minimum is for $U=0$
($S\rightarrow\infty$) so gauginos do not condense
(this is reflected in the runaway behaviour of the Wilsonian
action for $S$). For several condensing groups we find $U\neq 0$
and supersymmetry broken or not, depending on the case
\cite{ccm}.

\section{Linear vs Chiral Formalisms}

Here we report on the resolution of question (iv)
of section 2 \cite{bdqq1}\ :
perturbative $4D$ string theory has in its spectrum
a two-index
antisymmetric
tensor field $B_{\mu\nu}$.
Because it only has derivative
couplings, $B_{\mu\nu}$ is dual to a pseudoscalar field, the axion
$a$.
We can transform back and forth from the $B_{\mu\nu}$ and $a$
formulations as long as the corresponding shift
symmetries are preserved. It is known
that nonperturbative effects break the PQ symmetry of $a$
giving it a mass, then
the puzzle is: what happens to the stringy $B_{\mu\nu}$ field
in the presence of non-perturbative effects?
Is the duality symmetry also broken by those effects?
Is it then correct to forget about the $B_{\mu\nu}$ field, as
it is usually done, and work only with $a$? (Since,
unlike the axion,  $B_{\mu\nu}$ is
the field created by string  vertex operators).
The answer to these questions is very interesting:
duality symmetry is {\it not} broken by the
 nonperturbative
effects but the $B_{\mu\nu}$ field disappears from the propagating
spectrum! Its place is taken by a massive  $3$-index
antisymmetric tensor field $H_{\mu\nu\rho}$ dual to the
massive axion.

Here I will just sketch the main steps of the derivation
and refer the reader to \cite{bdqq1}\ for further details.
In $4D$ strings, the antisymmetric tensor belongs to a
linear superfield $L$ ($\overline{\cal DD}L=0$),
together with the dilaton
and the dilatino. For simplicity we only consider
the couplings of this field to gauge superfields in
global supersymmetry (the supergravity extension is straightforward),
 the most general action is then
the $D$-term of an arbitrary function $\Phi$,
${\cal L}_L = [ \Phi(\hat L)]_D$, with $\hat L\equiv
L-\Omega$ and $\Omega$ the Chern Simons superfield,
satisfying $\overline{\cal DD}\Omega=\WW$.

Since the gauginos appear in the lagrangian through the arbitrary
function $\Phi$, the analysis of gaugino condensation is far more
complicated in the linear case than in the chiral case. Furthermore,
the
Wilson action is not well defined in this case, because the field $L$
is
not gauge invariant, we cannot just integrate the gauge fields out
leaving an effective action for $L$ alone as we did for $S$.
Therefore we have to consider the 2PI action, and to
find it, we have to work in the first order formalism
where the gauge fields appear only through $\Tr\WW$ as
in the $S$ case. This will also allow us to perform  a duality
transformation and show that the $L$ and $S$ approaches are
equivalent.

 The duality transformation is obtained  by
starting with the first order system coupled to the external
current $J$:
\begin{eqnarray}
e^{ i{\cal W}(J)}
 &=& \int DV\, DS\, DY\, \exp
i\,
\int d^4x\; \left( \right.
\nonumber \\
 & & \left.
 {\cal L}(Y,S)
 +2\, \Re[J\Tr\WW]_F\right)
\end{eqnarray}

Where $V$ is the gauge superfield, $Y$ an arbitrary vector
superfield with the lagrangian  ${\cal L}(Y,S)=
\{\Phi(Y)\}_D
+\{S\overline{\cal DD}(Y +\Omega)\}_F$,
   and $S$ (the same $S$ of of the previous section!) starting life
as
a Lagrange multiplier chiral superfield.

Integrating out  $S$, implies
$\overline{\cal DD}(Y+\Omega)=0$ or $Y=L-\Omega\equiv\hat L$,
giving back the original theory. On the other hand
integrating first $Y$
gives the dual theory in terms of $S$ and $V$. This is the
situation above the condensation scale. Below condensation, however,
we have to integrate first the gauge fields, after that
we have the  same two options for getting the two dual theories,
the difference now is that the integration over $V$ breaks
the PQ symmetry (if there are at least
two condensing gauge groups)
and we are left with a duality without global symmetries.

To see this, we will concentrate on the
$2PI$ effective action $\Gamma(U,Y,S)$ obtained in the standard
way for $U\equiv\Avg{Tr\WW}$ \cite{bdqq2}.
The important result is that since $\cal W$ depends on
$S$ and $J$ only through the combination $S+J$, we can see
as in eq.~ (12) that
$\Gamma(U,S,Y)=US+\Xi(U,Y)$, where $\Xi(U,Y)$ is arbitrary,
therefore $S$ appears only
linearly in the path integral and its integration gives
again a $\delta$-function, but
imposing now $\overline{\cal{DD}}Y=-U$ instead of
the constraint $\overline{\cal DD}(Y+\Omega)=0$ above condensation
scale.
We can then see that there is no linear
multiplet implied by this new  constraint. This  is an indication
that
the $B_{\mu\nu}$ field is no longer in the spectrum.

The new propagating bosonic degrees of freedom in  $Y$ are,
 a scalar component, the dilaton, becoming
massive after  gaugino condensation and
a vector field $v^\mu$ dual to $a$, the pseudoscalar component of
$S$.
Instead of showing the details
of this  duality in components, I will describe the following
slightly
simplified toy model which has all the relevant properties:
$${\cal L}_{v^\mu,a} = -{1\over 2}v^\mu v_\mu-a \partial_\mu
v^\mu
-m^2 a^2 $$
If we solve for $v^\mu$ we obtain $v_\mu=-\partial_\mu a$,
substituting back we find
$${\cal L}_{a}={1\over 2}\partial^\mu a
\partial_\mu a -m^2 a^2$$
 describing the massive scalar
$a$. On the other hand, solving for $a$ we get
$a=-{1\over 2m^2}(\partial_\mu v^\mu)$ which gives
$${\cal L'}_{v^\mu}=
-{1\over 2}v^\mu v_\mu+{1\over 4m^2}(\partial_\mu v^\mu)^2.$$
 The lagrangian ${\cal L'}_{v^\mu}$
also describes a massive scalar given by the longitudinal, spin zero,
component of $v^\mu$.
 We can see that the only component that
has time derivatives is $v^0$, so the other three are
 auxiliary fields.
Notice that for $m=0$, we recover the standard duality
among a massless axion and  $B_{\mu\nu}$ field.
Therefore,
after the gaugino condensation process,
the original $B_{\mu \nu}$ field of the
linear multiplet is projected out of the spectrum in favour
of a massive scalar field corresponding to the
longitudinal component of  $v^\mu$ or
to the transverse component of the antisymmetric tensor
$H_{\mu\nu\rho}\equiv \epsilon_{\mu\nu\rho\sigma}v^\sigma$.
Thus solving the
puzzle of the axion mass in the two dual formulations.
Other interesting discussions of gaugino condensation in
the linear formalism can be found in \cite{others}.

\section{Scenarios for SUSY Breaking}

The results of the previous sections have shown us that
the general results extracted in the past years about
gaugino condensation in string models, in terms of the
field $S$,  are robust.
We have seen how gaugino condensation can in principle
lift the string vacuum degeneracy and break supersymmetry
at low energies (modulo de problems mentioned before).
But this is a very particular field theoretical mechanism
and it would be surprising that other nonperturbative effects
at the Planck scale could be completely irrelevant for these
issues. In general we should always consider the two types
of nonperturbative effects:stringy (at the Planck scale) and
field theoretical (like gaugino condensation). Four different
scenarios
can be considered depending on which class of mechanism solves
each of the two problems:lifting the vacuum degeneracy and breaking
supersymmetry.

 For breaking supersymmetry at low energies, we expect that
a field theoretical effect should be dominant in order
to generate the hierarchy of scales. 
We are then left with two preferred scenarios:
either the dominant nonperturbative effects are field theoretical,
solving both problems simultaneously, or there is a `two steps'
scenario
in which stringy effects dominate to lift  vacuum degeneracy
and field theory effects dominate to break supersymmetry.
The first scenario has been the only one considered so far,
the main reason is that we can control field theoretical
nonperturbative effects but not the stringy. In this scenario,
 independent of the particular mechanism, we have to face the
cosmological moduli problem.

In the two steps scenario
the dilaton and moduli fields are fixed at high energies
with a mass $\sim M_{Planck}$ thus avoiding the cosmological moduli
problem.
It is also reasonable to expect that Planck scale
effects can generate a potential for $S$ and $T$.
The problem resides in the implementaion of this scenario
\cite{biq},
mainly due
to our ignorance of nonperturbative string effects.

\subsection{S Duality}

To approach nonperturbative string effects
  we may use the conjectured $SL(2,Z)$
$S$-duality in $N=1$ effective lagrangians \cite{filq2}\ :
\be
S\rightarrow \frac{a\, S-i\, b}{i\, c\,S+d}, \qquad a\, d-b\, c=1.
\ee
Even though there is mounting evidence for this symmetry in
$N=4,2$ string backgrounds, it is not yet clear how
it will be extended to $N=1$ and if so most  probably the
lagrangian is not invariant under this symmetry
since it
usually  exchanges `electric' and `magnetic' degrees of freedom.
However, similar to the case of $T$ duality, if we restrict to
the part of the action that depends only on $S$,
(which is the relevant part when looking for vacuum configurations)
 this is expected to be
invariant under $S$ duality. Recall that if we do the same for
the classical action, the continuous $SL(2,R)$ transformation
is a symmetry of the truncated action, so the argument that quantum
effects break the continuous to the
discrete $S$ duality could actually make sense
in this case. As found in ref.~\cite{filq2}, the superpotential
should
be a modular form of weight $-1$ and can be written as:
\be
W(S)= \eta(S)^{-2}\, Q[j(S)]
\ee
where $Q$ is an arbitrary rational function of the absolute
modular invariant
function $j(S)$. Its arbitrariness forbids us to
extract concrete conclusions, but there are several general issues
worth mentioning.  Since the weight of $W(S)$ is negative, it
necessarily has poles \cite{cfilq}. If we further impose that
the scalar potential has to vanish at
 $S_R\rightarrow\infty$ (zero
string coupling)\cite{hm}\ there
should be poles at finite values of $S$ which may need
interpretation.
The functions $\eta(S)$ and $j(S)$ can be expressed as
infinite sums of $q\equiv e^{-2\pi S}$, thus encompassing
the expected nonperturbative instanton-like expansion.
The selfdual points $S=1,\exp{i\pi/6}$ are always extrema
of the potential and very often are minima. For those
points supersymmetry is unbroken, thus making the two
steps scenario very plausible at least for the $S$ field.

This way of fixing the vev of $S$ is much more elegant than
the racetrack scenario with several condensing gauge groups.
It is similar  to  the way  we understood the fixing of $T$.
A general question to be addressed to this scenario is that
usually the vev of $S$ is very close to $S_R\sim 1$ because
the nontrivial structure of the potentials is always close to the
selfdual points. This is far from the phenomenologically required
value where we want $4\pi/g^2\sim 25$. However, as emphasized
in \cite{lnn}\ the gauge coupling is $S$ only
at tree level,  it is expected to get nonperturbative corrections
and we may have a situation with $S_R=1$ but with a
larger value of $f(S)$ at the minimum leading to the
desired gauge coupling at the string scale.

Let us mention as an aside that the gaugino condensation
process can be made consistent with $S$-duality
\cite{hm,lnn,bg}. A way to do it is to write the
gaugino condensation superpotential $W\sim \exp{-\frac{3S}{c}}$
as the first term in an infinite expansion of the form
(15). Another approach is to try to {\it derive} the effective
superpotential from nonperturbative corrections to the
gauge kinetic function $f(S)$. The problem with this approach
is that we do not know how $f(S)$ should transform under $S$ duality
(we cannot forget the gauge fields as we did for finding  $W(S)$).
In ref.~\cite{lnn}, it was assumed that $f$ is invariant,
but then the gaugino condensation-induced superpotential
$W\sim \exp{-\frac{3f}{c}}$ would also be invariant
instead of a weight $-1$ form as required by $S$-duality.
 An extra factor $\eta(S)^{-2}Q[j(S)]$ has to be put in by hand
without justification, losing the connection with the
condensation process.

A probably better way  to derive an $S$ duality
invariant effective theory after gaugino condensation,
 may be to assume a
noninvariant $f(S)$ \cite{ilq}, after all that is
precisely what happens in $T$ duality
for which $f(T)\sim \log\eta(T)$. If for instance we take,
\be
f(S)=\frac{C}{\pi}\log\left\{\eta(S)\,
(j(S)-744)^{(C-12)/24C}\right\}
\ee
nonperturbatively (here $C$ is the Casimir of the
corresponding gauge group, see discussion below equation (5)),
 we can see that it has the right limit
for large $S$ ({\it ie} $f\rightarrow S$) and induces
a gaugino condensation superpotential $W(S)\sim \eta(S)^{-2}
(j(S)-744)^{(12-C)/12C}$ which has the right transformation
properties under $S$ duality and  reduces to the gaugino
condensation superpotential in the large $S$ limit.
The  noninvariance of $f(S)$ may probably
be related with $S$-duality anomalies \cite{ilq}\ as it happened
in the $T$ duality case.
A problem with this approach is that if we are considering
nonperturbative corrections to the $f$ function, we should also
include those corrections for $W$ and $K$. This may
diminish the importance of the gaugino condensation-induced
superpotential
above, because it would be just an extra contribution to
the original nonperturbative superpotential which we do not know.
There may still be situations, as argued in \cite{bd}, for which
gaugino condensation superpotentials could nevertheless
be dominant.

\subsection{Two Steps Scenario}

In the two steps scenario, after we have fixed the vev
of the moduli by stringy effects, it remains the question of
how supersymmetry is broken at low energies. Notice that we
would be left with the situation present before the advent
of string theory in which the gauge coupling is
 field {\it independent}. In that
case we know from Witten's index that gaugino condensation cannot
break global supersymmetry. Since there are no `moduli'
fields with large vev's, the supergravity correction should
be negligible because we are working at energies much smaller
than $M_{Planck}$.

 In fact we can perform a calculation by
setting $S$ to a constant in eq.~(12),
 it is straightforward to show that supersymmetry
is still unbroken in that case \cite{biq}, as expected. A more
general
way to see this is
computing explicitly the $1/M_{Planck}$ correction to a
global supersymmetric solution $W_\phi=0$, and see that it
coincides with the solution of
$W_\phi+WK_\phi/M_p^2=0$ which  is always a
supersymmetric extremum of
the supergravity scalar potential.

As mentioned in section 2, there seems to be however a counterexample
in the literature.
In ref.~\cite{peter}  a modification of the
K\"ahler potential (12) was considered:
\be
e^{-K/3}=1-a\, \left(UU^*\right)^{1/3}-b\, \left(UU^*\right)
\ee
with the same superpotential. For $a=-9b$
supersymmetry was found to be  broken with vanishing cosmological
constant.
But also for this choice of parameters
the global limit is such that $K_{UU^*}$ vanishes, and so the kinetic
energy
for $U$. This makes the corresponding
 minimum in the global case ill defined, since there may be other
 nonconstant field configurations with vanishing energy.
This is then not a counterexample, because the
global theory is not well defined in the minimum.
In any case, in our general analysis, there are no such extra
corrections to the K\"ahler potential for $U$.

We are then left with a situation that if global supersymmetry
is unbroken, we cannot break local supersymmetry, unless
there are moduli like fields.
This can bring us further back to the past and reconsider models
with dynamical
breaking of global supersymmetry
(for a recent discussion with new insights see \cite{dnn}\ ).

\section{Conclusions}

\begin{description}

\item[(i)] Gaugino condensation provides a simple example
of how supersymmetry can be broken dynamically with
partial succes. Some of the problems may be solved
after having better control of the supergravity
lagrangian. In particular, in the single hidden sector
group case we have seen that the gauginos do not
condense, but this situation may be changed after perturbative and
nonperturbative corrections to the K\"ahler potential are considered
\cite{bd}. The cosmological problems may be
more generic, however.

\item[(ii)] The gaugino condensation process
is also an interesting laboratory to test
nonperturbative properties of string and field theories. In
particular
duality symmetries survive this simple,
but nontrivial, nonperturbative test.

\item[(iii)]The different approaches to describe the effective
theory underlying the condensation process correspond
simply to the use of the Wilson or 2PI effective actions, therefore
there is
a well defined relation among them.
Even though the Wilson action is usually simpler to work with,
the 2PI action is more suitable to follow the condensation process,
it also is the only one that could be used to describe
the condensation of gauginos in the `linear formalism'.
The Wilson action cannot be used without previously identifying the
low energy degrees of freedom.
We needed the 2PI action to find out that the
axion degree of freedom is represented by a massive
 $H_{\mu\nu\rho}$ tensor.

\item[(iv)] The linear and chiral descriptions are equivalent,
even in the absence of PQ symmetries.
Which formulation is more convenient depends on the situation.
In the linear description, the stringy $B_{\mu\nu}$
field is replaced by the massive $H_{\mu\nu\rho}$ field.
We believe, this will also be the case in
more general nonperturbative effects.
We may conjecture that this result could be related with the claims
that `stringy' nonperturbative effects  are
not well described by strings but better by membranes, which
couple naturally to $H_{\mu\nu\rho}$ or five-branes, which
provide the 10D origin of the field $S$. A (massless) field
$H_{\mu\nu\rho}$ also
 appears naturally in 11D supergravity.

\item[(v)] There is not a compelling
 scenario for supersymmetry breaking
and the field remains open, but we have a much
better perspective on the relevant issues now. The
nonrenormalizable hidden sector models of which the
gaugino condensation is
a particular case, may need a convincing  solution of the
cosmological moduli problem to still be considered
viable. Hopefully, this will lead to interesting feedback
between cosmology and string theory \cite{bms}.
Furthermore, the recent progress in understanding
supersymmetric gauge theories can be of much use for
reconsidering gaugino condensation with hidden matter,
the discussion in the string literature is far from complete.
The understanding of models with chiral
matter could also provide new insights to
global supersymmetry
breaking, relevant to the
two steps scenario mentioned above. In any case the techniques
found to be useful in the simplest gaugino condensation
approach discussed here, will certainly help in understanding those
  more
complicated models.
\end{description}

I thank the organizers for the invitation to
participate in such an exciting conference.


\begin{thebibliography}{19}
\bibitem{witten}{E. Witten, Nucl. Phys. B188 (1981) 513; Nucl. Phys.
B202 (1982) 253}
\bibitem{vy}{G. Veneziano, S. Yankielowicz, Phys. Lett. 113B (1982)
231.
\bibitem{amati}{D. Amati, K. Konishi, Y.Meurice, G.C. Rossi,
G. Veneziano, Phys. Rep. 162 (1988) 169.}
\bibitem{peter}{H.P. Nilles, Phys. Lett. 115B (1982) 193.}
\bibitem{fgn}{S. Ferrara, L. Girardello, H.P. Nilles, Phys. Lett.
125B (1983) 457.}
\bibitem{drsw}{M. Dine, R. Rohm, N. Seiberg, E. Witten,
Phys. Lett. 156B (1985) 55.}
\bibitem{din}{J.-P. Derendinger, L.E. Ib\'a\~nez, H.P. Nilles,
Phys. Lett. 155B (1985) 65.}
\bibitem{filq}{A. Font, L.E. Ib\'a\~nez, D. L\"ust, F.
Quevedo, Phys. Lett. 245B (1990) 401;
S. Ferrara, N. Magnoli, T. Taylor, G. Veneziano, Phys. Lett.
245B (1990) 409; H.P. Nilles, M. Olechowski, Phys. Lett. 248B (1990)
268;
P. Binetruy, M.K. Gaillard, Phys. Lett. 253B (1991) 119.}
\bibitem{dkl}{L. Dixon, V. Kaplunovsky, J. Louis, Nucl. Phys. B329
(1990) 27.}
\bibitem{krasnikov}{N. Krasnikov, Phys. Lett. 193B (1987) 37;
L. Dixon, in {\it The Rice Meeting}, B. Bonner, H. Miettinen, eds,
World Scientific (1990); J.A. Casas, Z. Lalak, C. Mu\~noz, G.G. Ross,
Nucl. Phys. B347 (1990) 243.}
\bibitem{lt}{D. L\"ust, T. Taylor, Phys. Lett. 253B (1991) 335.}
\bibitem{ccm}{B. de Carlos, J.A. Casas, C. Mu\~noz, Nucl. Phys. B399
(1993) 623.}
\bibitem{bs}{R. Brustein, P. Steinhardt, Phys. Lett. B302 (1993)
196.}
\bibitem{modprob}{T. Banks, D. Kaplan, A. Nelson, Phys. Rev. D49
(1994) 779;
B. de Carlos, J.A. Casas, F. Quevedo, E. Roulet, Phys. Lett. B318
(1993) 447.}
\bibitem{mr}{A. de la Macorra, G.G. Ross, Nucl. Phys. B404 (1993)
321;
R. Peschanski, C. Savoy, hep-ph/9504243.}
\bibitem{kp}{C. Kounnas, M. Porrati, Phys. Lett. 191B (1987) 91.}
\bibitem{kl}{V. Kaplunovsky and J.Louis, Nucl. Phys. B422 (1994) 57.}
\bibitem{bdqq2}{C.P. Burgess, J.-P. Derendinger, F. Quevedo,
M. Quir\'os,  hep-th/9505171.}
\bibitem{cfgvp}{E. Cremmer, S. Ferrara, L. Girardello,
A. van Proeyen, Nucl. Phys. B212 (1983) 413.}
\bibitem{dfkz}{J.-P. Derendinger, S. Ferrara, C. Kounnas
F. Zwirner, Nucl. Phys. B372 (1992) 145;Phys. Lett. 271B (1991) 307.}
\bibitem{bdqq1}{C.P. Burgess, J.-P. Derendinger, F. Quevedo and
M. Quir\'os, Phys. Lett. 348B (1995) 428.}
\bibitem{others}{
J.-P. Derendinger, F. Quevedo, M. Quir\'os,
Nucl. Phys. B428 (1994) 282;I. Gaida, D. L\"ust, Int. J. Mod. Phys.
A10 (1995) 2769;
hep-th/9510022; P. Binetruy, M.K. Gaillard, T.
Taylor, hep-th/9504143.}
\bibitem{biq}{C.P.  Burgess, L.E. Ib\'a\~nez, F. Quevedo,
unpublished.}
\bibitem{filq2}{A. Font, L.E. Ib\'a\~nez, D. L\"ust, F.
Quevedo, Phys. Lett. 249B (1990) 35.}
\bibitem{cfilq}{M. Cveti\v c, A. Font, L.E. Ib\'a\~nez, D. L\"ust,
F. Quevedo, Nucl. Phys. B361 (1991) 194.}
\bibitem{lnn}{Z. Lalak, A. Niemeyer, H.P. Nilles, Phys. Lett.
349B (1995) 99; hep-th/9503170.}
\bibitem{hm}{J. Horne, G. Moore, Nucl. Phys. B432 (1994) 109.}
\bibitem{bg}{P. Binetruy, M.K. Gaillard, hep-th/9506207.}
\bibitem{ilq}{L.E. Ib\'a\~nez, D. L\"ust, F. Quevedo, unpublished.}
\bibitem{bd}{T. Banks, M. Dine, Phys. Rev. D50 (1994) 7454.}
\bibitem{dnn}{M. Dine, A. Nelson, Y.Nir, Y. Shirman,
hep-ph/9507378.}
\bibitem{bms}{T. Banks, M. Berkooz, G. Moore, S. Shenker, P.
Steinhardt,
Phys. Rev. D52 (1995) 3548; T. Banks, M. Berkooz, P. Steinhardt,
Phys. Rev. D52 (1995) 705; A. de la Macorra, hep-ph/9501250.}}

\end{thebibliography}
\end{document}